\documentclass[useAMS,usenatbib]{mn2e}
\usepackage{epsfig}

\title[]{
Radial mixing in the outer Milky Way disk caused by an orbiting 
satellite 
}
\author[]{Alice C. Quillen$^1$, Ivan Minchev$^2$, Joss Bland-Hawthorn$^3$, 
\& Misha Haywood$^4$\\
{$^1$ Department of Physics and Astronomy, University of Rochester, Rochester, NY 14627, USA; aquillen@pas.rochester.edu} \\
{$^2$ Universit\'e de Strasbourg, CNRS, Observatoire Astronomique, 11 rue de l'Universit\'e, 67000 Strasbourg, France; minchev@astro.u-strasbg.fr} \\
{$^3$ Sydney Institute of Astronomy, School of Physics, University of Sydney, NSW 2006, Australia;  jbh@physics.usyd.edu.au} \\
{$^4$ GEPI, Observatoire de Paris, F-92195 Meudon Cedex, France}
}

\begin{document}

\label{firstpage}
\maketitle

\begin{abstract}
Using test particle simulations we examine the structure
of the outer Galactic disk as it is perturbed by
a satellite in a tight eccentric orbit about the Galaxy.
A satellite of mass a few times $10^9 M_\odot$
can heat the outer Galactic disk, excite spiral
structure and a warp and induce streams in the
velocity distribution.  
We examine particle eccentricity versus the change
in mean radius between initial and current orbits. 
Correlations between these quantities are
reduced after a few satellite pericenter passages. 
Stars born in the outer galaxy
can be moved in radius from their birth positions
and be placed in low eccentricity orbits inside their birth
radii.  We propose that mergers and perturbations 
from satellite galaxies and subhalos can induce radial
mixing in the stellar metallicity distribution.

\end{abstract}

\section{Introduction}

There is evidence for past and ongoing accretion of small objects
by the Milky Way, the most dramatic one
being the disrupted Sgr dwarf galaxy \citep{ibata94}.
The outer disk of
the Galaxy may be constantly perturbed by small satellites 
or structure in the galactic halo which
can contribute to the stellar halo population \citep{helmi08}.
Perturbations from satellite galaxies or merging subhalos
with mass of order $10^{10} M_\odot$
can cause ring-like structures in the Galactic
plane \citep{tutukov06,younger08,kazantzidis08} similar to the 
Monoceros stellar stream \citep{newberg02,juric08}.
Cosmological simulations suggest that such
events are not uncommon. Our
galaxy could have experienced pericenter
crossings within 20kpc of the Galactic
center with about 6 separate subhalos of this size
since a redshift of 1 \citep{kazantzidis08} (also see
\citealt{purcell07,purcell09}).
Mergers and orbiting satellite subhalos can 
induce a warp, thicken the Galactic disk 
and leave behind stellar streams 
\citep{bekki03,helmi03,meza05,penarrubia05,purcell07,hopkins08,kazantzidis08,villalobos08,purcell09}.

The metallicities of stars in the solar neighborhood
exhibit a large scatter that is not strongly dependent
on their age or orbital eccentricity \citep{edvardsson93,haywood08}.
The lack of strong correlations in these properties
can be explained if stars migrate radially
from their birthplace \citep{haywood08,schoenrich09}.
Radial migration can be caused by resonances
associated with transient spiral structure \citep{sellwood01,roskar08}.
Here we explore the possibility that radial migration
can be induced tidally by an orbiting satellite galaxy.

\section{Test Particle Simulations}

The numerical simulations integrate test particle trajectories
in a gravitational potential in cylindrical coordinates
with a flat rotation curve
\begin{equation}
\Phi(r,z) = {v_c^2 \over 2} \log (r^2 + z^2/q^2 + a^2)
\end{equation}
where $q<1$ corresponds to an oblate system and
$a$ is a smoothing length or core radius.
Disk particles are initially chosen with radial 
distribution $n(r) \propto r^{-1}$.
Azimuthal and epicylic angles are randomly chosen.
Particle eccentricities and inclinations
are initially chosen from a Rayleigh distribution
so as to create 
a Gaussian distribution in both radial and vertical velocities.

A single massive particle is placed at apocenter to mimic the
effect of the dwarf satellite and integrated in the same
potential as the disk particles.
The disk particles feel the perturbations from the dwarf
galaxy, though not each other.   
The dwarf galaxy is put on an orbit with eccentricity
and orbital period similar to those
used to predict the location and motions of the Sgr
dwarf debris \citep{johnston99,law05}.
We chose an orbit for the dwarf galaxy similar
to that of Sgr dwarf because this orbit is well constrained
by observations and because its pericenter is small enough
that a more massive dwarf galaxy
in this orbit would have strongly perturbed the Milky Way disk.
The force from the satellite is softened at short
distances  so that spurious large velocities are not induced
when it passes through the Galactic disk.
The test particles are integrated for
four radial oscillations periods of the dwarf galaxy orbit.
Thus the disk feels four close pericenter passages.
The gravitational potential of the Galaxy is not varied during
the simulation, thus there is no indirect perturbation
that should have been caused by variation in the position
of the center of the galaxy with respect to the center of mass frame.
As these are test particle simulations, the self-gravity 
of the disk and response of the halo to the satellite
are neglected.

We work in units of the Sun's radius from
the Galactic center ($R_0 \approx 8.5$~kpc) and the rotational
velocity for a particle in a circular orbit at $R_0$ 
or $v_0 \approx$ 220~km/s.
Masses are given in units of 
$M_0 = {v_0^2 R_0 \over G} \approx 10^{11} M_\odot$.
Timescales are given in units of the rotation period at $R_0$ or 
$P_0 \approx $ 240~Myr.
Initial disk particles are chosen between 0.5 and 2.5 $R_0$.
The softening length for the satellite is 0.1 $R_0$
and that for the rotation curve is $a =0.1 R_0$ and within
the radii of disk particles simulated.
The simulations are run with code described by \cite{bagley09}
but lacking a bar perturbation.  To compute spatial distributions,
we used $10^5$ particles.  To compute
mean velocity components, mean disk height and
angular momentum and energy distributions we used simulations
with $10^6$ particles.
We used simulations of $10^7$ disk particles
to compute velocity distributions of stars in
the neighborhood of a point in
the disk.  This allows us to search for fine structure 
in the velocity distribution of stars near a specific
location in the disk (like the solar neighborhood).

While the current mass of Sgr dwarf galaxy $\sim 5 \times 10^8 M_\odot$
is not massive enough to strongly perturb the Galactic disk,
this galaxy could have been a few times $10^9 M_\odot$ in
the past (e.g., see discussion
by \citealt{law05,zhao04}) and so similar in mass to the subhalos considered
by \citet{kazantzidis08}.
In the context of a CDM hierarchical galaxy
formation paradigm, \citet{kazantzidis08} show that it is not
uncommon for a subhalos of sizescale $\sim 10^{10} M_\odot$
to merge with a Milky Way sized galaxy. 
We consider satellites on tight orbits with pericenters 
approximately $1.3 R_0$ and masses 0.03 to 0.1 $M_0$ 
($3$ to $10 \times 10^9 M_\odot$).
Our satellites have pericenter and period similar to the shorter values
estimated for Sgr galaxy.
In our units the estimated
peri and apocenter Galactocentric radii of Sgr galaxy's orbit
are 1.3--2.0 and 6.5--7.0 \citep{law05}.  
Our simulation parameters are listed in Table \ref{tab:tab1}.
The galaxy halo potential is chosen to be mildly oblate with $q=0.95$.
We chose a non-spherical potential so as to allow
mild variations between pericenter galactocentric radii. 
Variations in the orbit would have been present 
if dynamical friction were present 
or if the Milky Way gravitational potential were allowed
to respond to the perturbation.

\subsection{Morphology}

\begin{figure}
\begin{center}$
\begin{array}{cc}
\includegraphics[angle=0,width=3.5in]{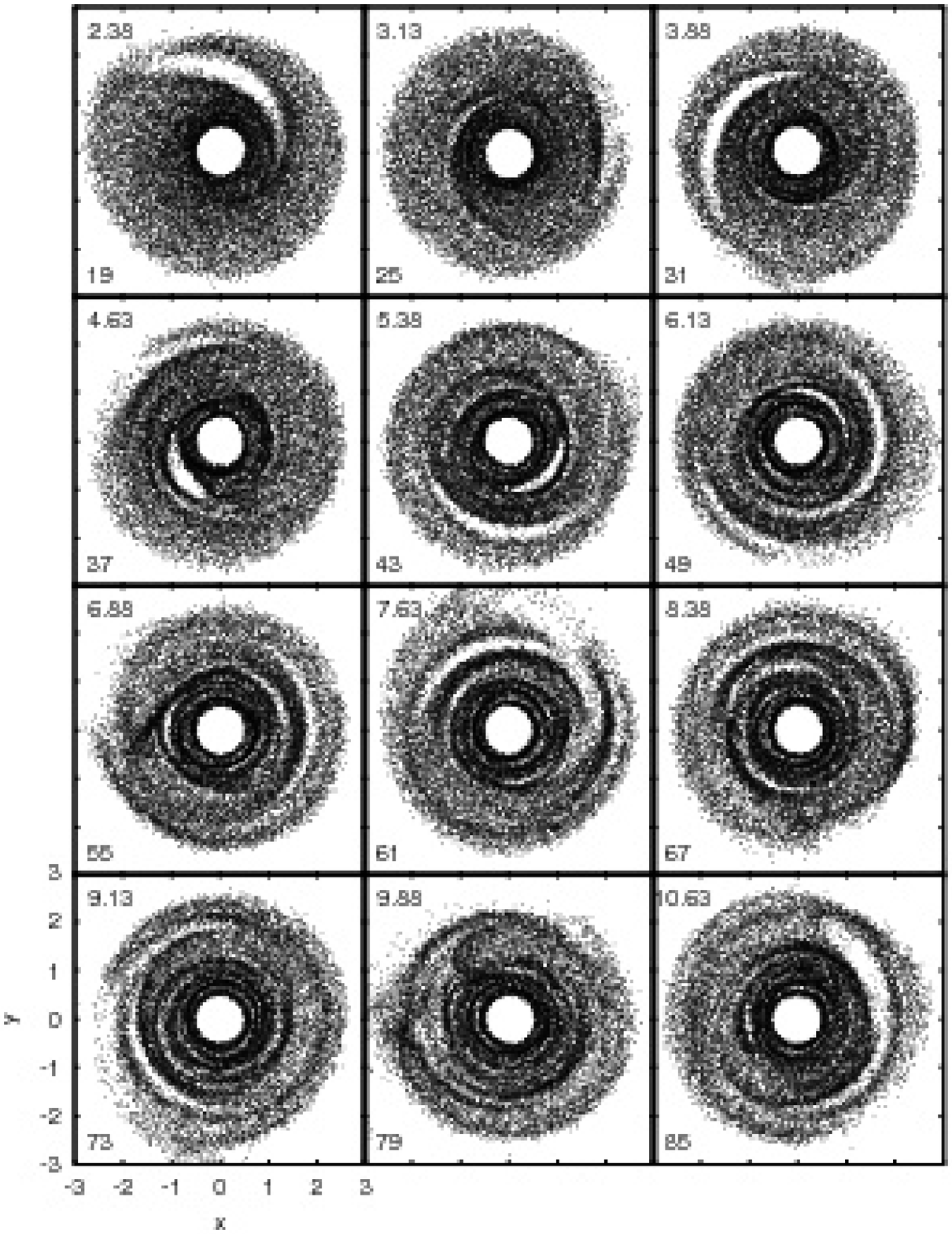} &
\includegraphics[angle=0,width=3.5in]{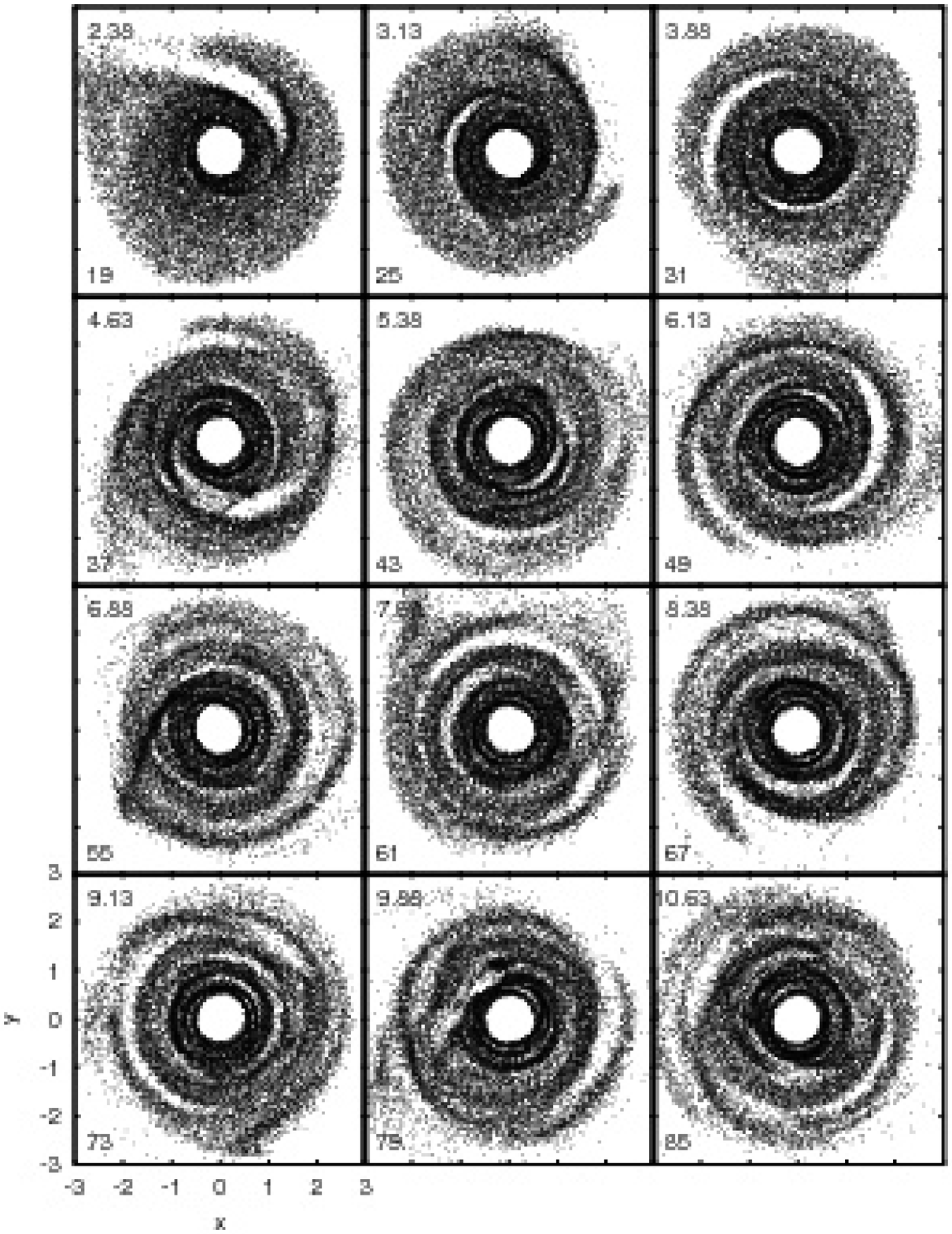}
\end{array}$
\end{center}
\caption{
a) Structure in the plane of the galaxy for the A1 
test particle simulation.
Spiral structure and lopsidedness in the disk has been
excited by a $6 \times 10^{9} M_\odot$ satellite galaxy in
an polar orbit approximating that of the Sgr dwarf galaxy.
The simulation has parameters listed in Table \ref{tab:tab1}.
The image is oriented so that galactic rotation is clockwise.
Numbers on the upper left in each panel show time since
simulation start in units of the rotation period at the Sun.
Number on the lower left in each panel show the output number
as positions were output every 1/8 period.
Pericenters for the satellite occur at output stages 12, 33, 54 and 75.
The outer disk exhibits spiral structure, a warp and is lopsided.
These structures are caused by perturbations from the satellite.
b) same as a) except for simulation A2 with satellite on
an inclined prograde rather than polar orbit.
Perturbations on the disk are stronger for this simulation.
\label{fig:xyh}
}
\end{figure}

In Figure \ref{fig:xyh}a we show the morphology
in the plane of the Galaxy as a function of time from
the simulation A1 with parameters listed in Table \ref{tab:tab1} and
satellite on a polar orbit.
Images are shown separated in time by 3/4 of $P_0$,
the orbital period at the Sun.
In Figure \ref{fig:xyh} the galaxy is oriented so that rotation is clockwise.

The simulation shown in Figure \ref{fig:xyh}a
has satellite orbit resembling that of Sgr dwarf.
The orbit is oriented so that is it polar or lying in the $y,z$ plane.
The first pericenter passage is experienced 
1.5 orbital periods (at the Sun) from the beginning 
of the simulation.
The first pericenter occurs below the Galactic plane.
The second and third pericenter passages occur closer
to the Galactic plane.  
The pericenters close approaches and passages through the 
plane of the Galaxy induce 
velocity impulses in the stars in Galactic disk.  
Following the second pericenter spiral structure
is induced in the disk with structure similar
to that seen in simulations of flybys \citep{tutukov06,younger08}.
The disk also exhibits some lopsidedness though not as much
as displayed by the atomic hydrogen distribution 
in the Milky Way's outer disk \citep{levine06a}.

Multiple close passages cause stronger spiral structure
than a single flyby.
This is not surprising as the tidal
force during pericenter lasts longer when the orbit is bound 
rather than parabolic, and 
the disk cannot completely dynamically relax between pericenters.
The disk receives multiple perturbations from each close approach 
which can add constructively or destructively to particle
eccentricity or inclination as
the timescale between pericenters, about 3 $P_0$, is comparable
to the orbital period  at 2 $R_0$ (or 2 $P_0$).

Inspection of Figure \ref{fig:xyh} shows that
particle orbits can intersect other orbits.  Stellar
orbits can cross each other but gas clouds would collide.
As we see orbits that intersect each other,
the gas distribution may differ from
the stellar distribution.  It is interesting to place
this in perspective with the outer Galaxy. 
There is no clear structure in the HI
distribution (as seen in the maps
by \citealt{levine06a})
in the Milky Way associated with the Monoceros stream 
(as seen in the maps by \citealt{juric08})
even though there is HI gas detected in the same region. 

Figure \ref{fig:xyh}b is similar to Figure \ref{fig:xyh}a
except showing a the disk perturbed by a satellite
on a prograde inclined orbit (simulation A2).  The orbit's spin axis
is 30$^\circ$ from the Galactic pole. 
We see that the perturbations on the disk 
are stronger for the inclined orbit.  The outer disk
is more strongly scattered and the induced spiral structure
is stronger. 

\begin{figure}
\begin{center}$
\begin{array}{cc}
\includegraphics[angle=0,width=2.7in]{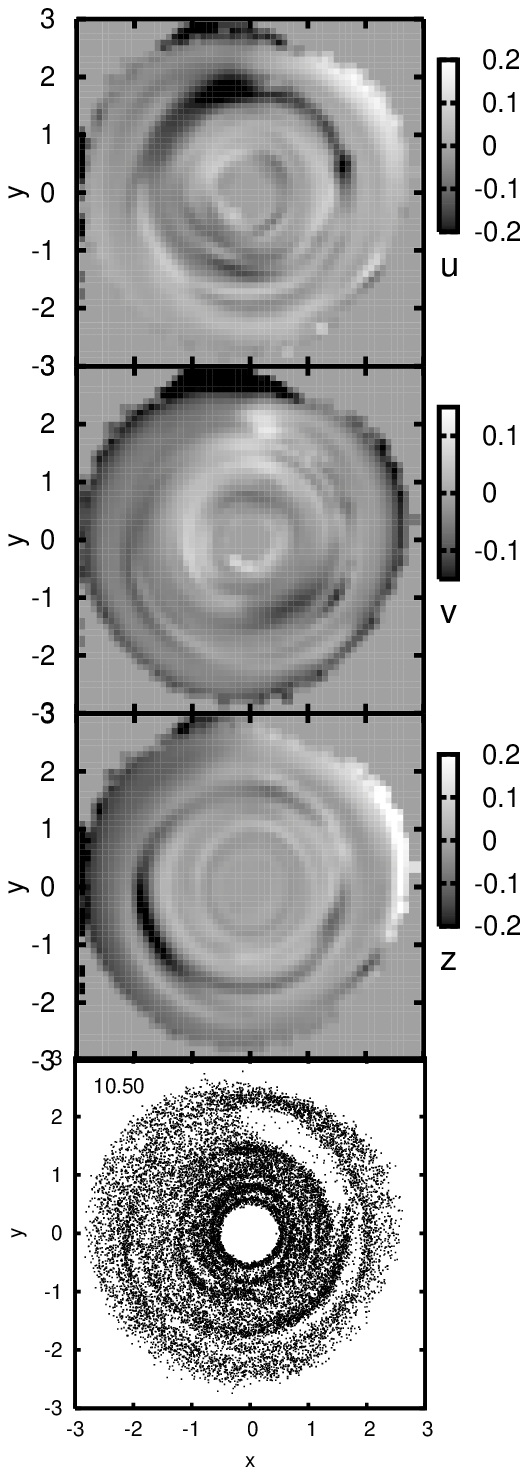} &
\includegraphics[angle=0,width=2.7in]{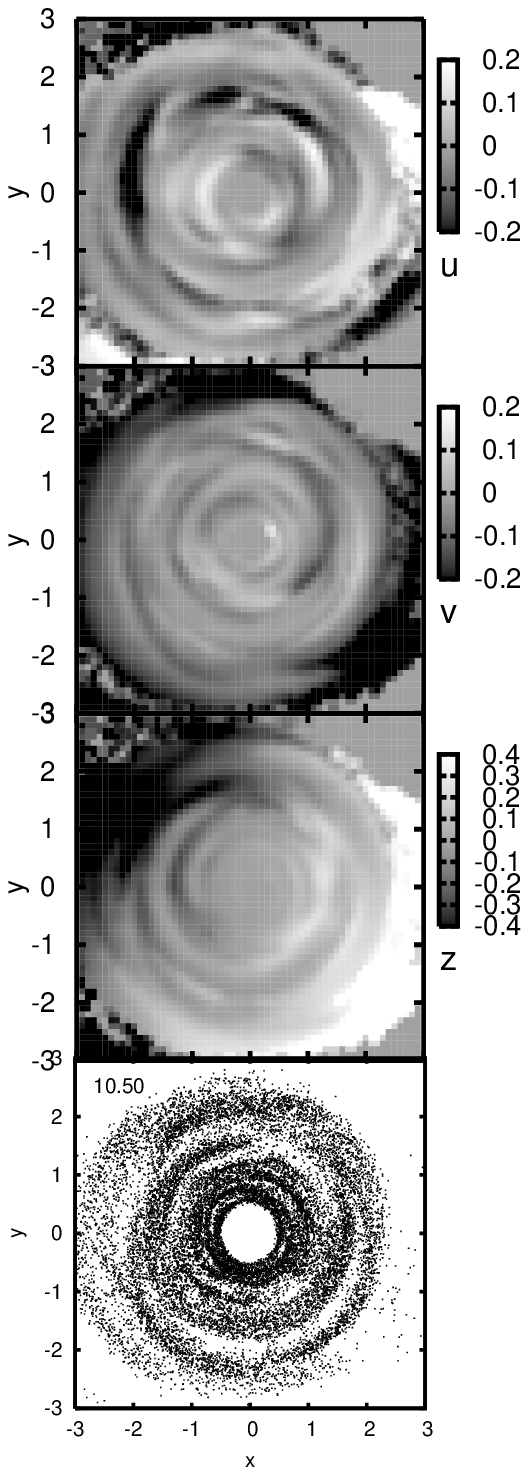}
\end{array}$
\end{center}
\caption{From top to bottom are shown 
the mean radial velocity, $u$ (with positive $u$ away from
the Galactic center), the mean tangential velocity
subtracted by the circular velocity, $v$, the 
mean height above or below the Galactic plane, $z$, 
and the disk particles projected into the  plane of the galaxy.
These distributions are shown when the satellite is has
just passed its fourth pericenter. 
or at 10.5  orbital periods ($P_0$) after start of simulation.
The disk lopsidedness and warp have been induced by the satellite.
a) (on left) For Simulation A1 with satellite on a polar orbit.
b) (on right) For Simulation A2 with satellite on an inclined orbit.
The velocity scale is in units of the circular velocity at $R_0$.
The $z$ scale is in units of $R_0$.
\label{fig:sag}
}
\end{figure}

After the last pericenter passage,
the mean height of the disk above
the Galactic plane, radial and tangential velocity components 
(in Galactocentric coordinates) are shown in Figure \ref{fig:sag} 
for both simulations.
For these plots the radial velocity component is positive in
the direction away from the Galactic center.
The velocity of a particle in a circular orbit
has been subtracted from the tangential velocity component.

As have other works (e.g., \citealt{tutukov06,younger08,kazantzidis08}) 
we find that the satellite induces
spiral structure in the outer disk that is similar 
to that in Monoceros stream.  
The tangential velocity component is not expected to be high
(second panels from top in Figure \ref{fig:sag})
but there should be moderate radial velocity variations 
(top panels in Figure \ref{fig:sag}).
A comparison between the mean velocity components, the density 
distribution and the mean height (shown
in the second panels from bottom in Figure \ref{fig:sag}) 
suggests that there could be correlations
between these quantities in the outer
Milky Way, though as we have neglected self-gravity
in the disk, we should be cautious with this implication.

We see from the second panel from the bottom in Figure \ref{fig:sag} 
that a warp has been excited by the passages of the satellite.
The warp is complex and not well described by a single
Fourier component as is true for the outer Milky Way \citep{levine06b}.  
However, its amplitude for the polar orbit simulation is smaller than
that observed in the outer Milky Way even though the satellite,
$6 \times 10^{9} M_\odot$  exceeds that estimated
for the Sgr dwarf galaxy by a factor of 10.  
The outer Milky Way extends out of the
plane 4-6 kpc which would be $z\sim 0.5$ in units of $R_0$.
From Figure \ref{fig:sag} we see that the mean 
height reaches only $z \sim 0.2$.  A similar
mass satellite on a tight orbit but inclined
to the disk does induce a warp large enough
to account for features seen in the outer galaxy (as
shown in Figure \ref{fig:sag}b).  
There is currently no evidence for a coherent satellite
with mass as large as we are simulating that could
account for the morphology of the outer galaxy. 
As explored by \citet{weinberg98,weinberg06} with the LMC 
resonant perturbations could be important and a self-gravitating disk
would be needed to more accurately simulate the coupling, or the response
of the Milky Way halo is important 
and a live halo would be needed to simulate the effect. 
Alternatively objects as massive as simulated here, possibly
merging with the outer Galaxy, are needed to explain the 
Galactic warp. 

\subsection{Velocity distribution at a single position}

\begin{figure}
\includegraphics[angle=0,width=5.0in]{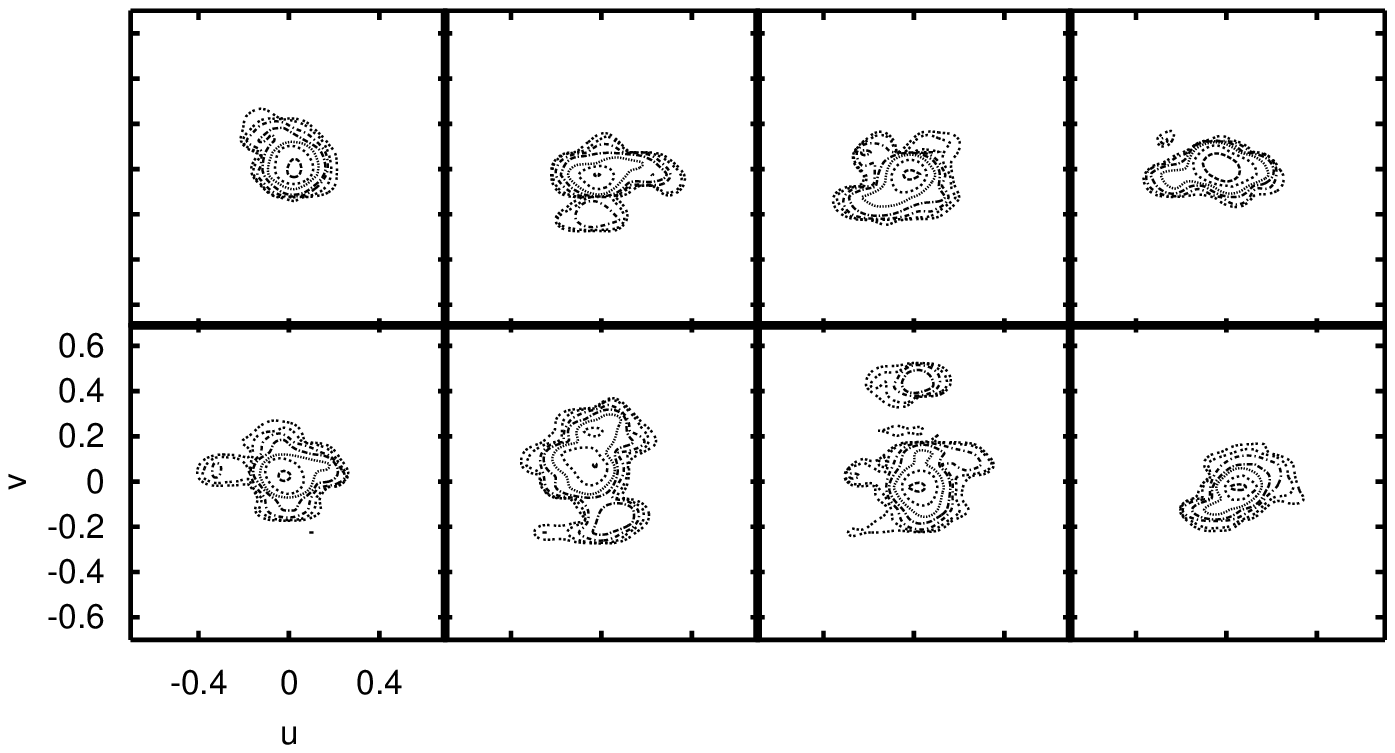}
\includegraphics[angle=0,width=5.0in]{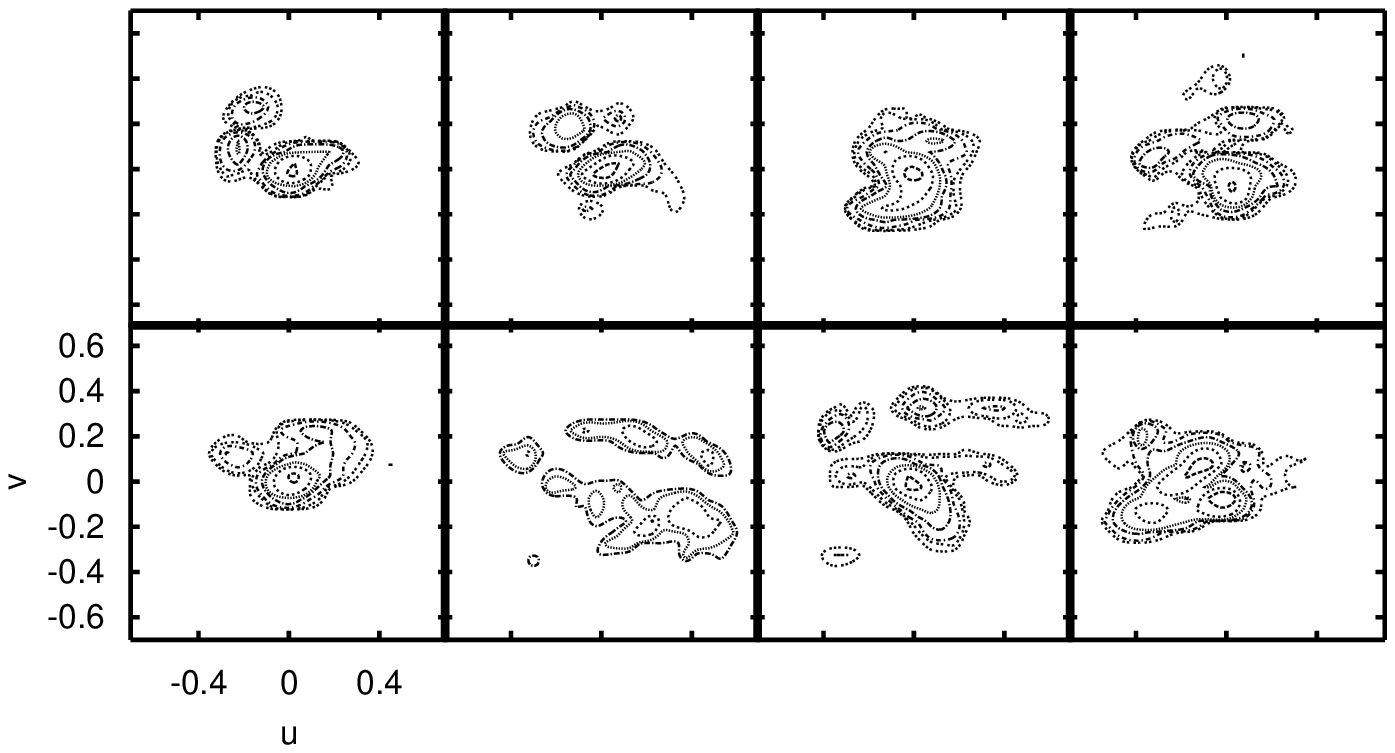}
\caption{The velocity distribution 
in a particular region of the disk (like the solar neighborhood).
a) For simulation A1.
Velocity distributions
are shown for stars within 0.05 $R_0$ of the position of
the Sun, at (0,1) in Figure \ref{fig:xyh}. 
The first panel (on top left) is 7.5 orbital periods
after the beginning of the simulation. 
Each panel is separated in time by a half orbital period.
Contours are logarithmically spaced with the lowest contour at 1 particle
per $0.05\times 0.05$ square velocity bin.
There are 2 contours within each factor of 10.  10 million particles
were integrated to construct the velocity distribution
in the small volume near the Sun. 
We find that clumps in the solar neighborhood velocity distribution
can be induced by tidal effects associated with
perturbations from a moderate mass (here $6 \times 10^9 M_\odot$)
satellite in a tight eccentric orbit around the Galaxy.
Higher velocity streams are found at positive $v$ and correspond
to stars coming into the solar neighborhood from the outer
galaxy.
As the satellite is not allowed to decay, disrupt or
merge, the Milky Way
disk is repeatedly perturbed.  The disk will not relax to
a smooth distribution until the perturbations cease.
b) For simulation A2. 
The satellite in an inclined orbit causes higher velocity streams
than when the satellite orbit is polar as for simulation A1.
\label{fig:uv}
}
\end{figure}

We consider whether the perturbations caused by
the satellite would  cause structure in the
velocity distribution in a particular location
such as the solar neighborhood.
Figure \ref{fig:uv} shows the velocity distribution computed
in a region centered at the solar position from stars
that are within a distance of $0.05 R_0$ or $\sim 400$~pc.
We find that spiral arms induced by the satellite can
increase stellar eccentricities sufficiently 
that streams can be seen in the velocity
distribution.  Large velocity streams are not well populated 
but streams at about 40 km/s (at $u$ or $v \approx \pm 0.2$ 
from the local standard of rest in units of $V_0$)
are seen in the simulations, particularly at positive $v$.
Stars at positive $v$ correspond to those from the outer galaxy with
eccentricity high enough that they can be seen in the solar
neighborhood.  These stars would correspond to a disk
population that has lower metallicity than the thin
disk population present in the solar neighborhood because
they were born in the outer galaxy.  Such a population
with origin in the outer disk has been
postulated by \citet{haywood08}. 

Small population streams of stars at high velocity in the
solar neighborhood have primarily been 
interpreted as remnants from disrupted satellites
(e.g., \citealt{navarro04,helmi06}). Here we
find that they
could also be comprised of disk stars but associated
with recent tidal perturbations to the disk caused
by moderate mass satellites or mergers.

The relaxation or sheering timescale (in
rotation periods) for epicyclic motions is fairly short.
After perturbations cease, structure in
the velocity distribution will become less and less
prominent.  
\citet{minchev09} shows that streams could
last a few Gyrs but would get closer and closer together
at later times.  
In our simulations 1 Gyr corresponds approximately to 4 orbital periods.  
In our simulations because perturbations recur,
the disk would never be relaxed and multiple streams
would always be present.   If the satellite were
allowed to decay via dynamical friction, be disrupted and merge
with the Galaxy then we expect that the streams
would become closer together finally dissolving into a diffuse
distribution a few Gyrs afterward the merger.

\subsection{Angular momentum and eccentricity changes}

After the satellite has disrupted or merged with
the Galaxy we expect spiral structure induced
in the disk will wind up in 5-10 rotation 
periods.
Compact stellar streams
such as seen in Figure \ref{fig:uv} will wrap into
a diffuse smooth distribution via dynamical relaxation 
(e.g., \citealt{minchev09}).  However stars excited to
higher eccentricities are likely to remain
at higher eccentricity.  We now look at angular
momentum and eccentricity changes in the disk
induced by the satellite.

For 3 different ranges of initial radii  and at different
times we compute
$dL$,  the angular momentum change and  $e$, the particle
eccentricity.  The angular momentum
change is computed by subtracting the current one from
the initial one.  The eccentricity is estimated
as $e \sim \sqrt{(u^2 + 2 v^2)/2}$ appropriate for
a flat rotation curve \citep{arifyanto06}.

\begin{figure}
\includegraphics[angle=0,width=5.0in]{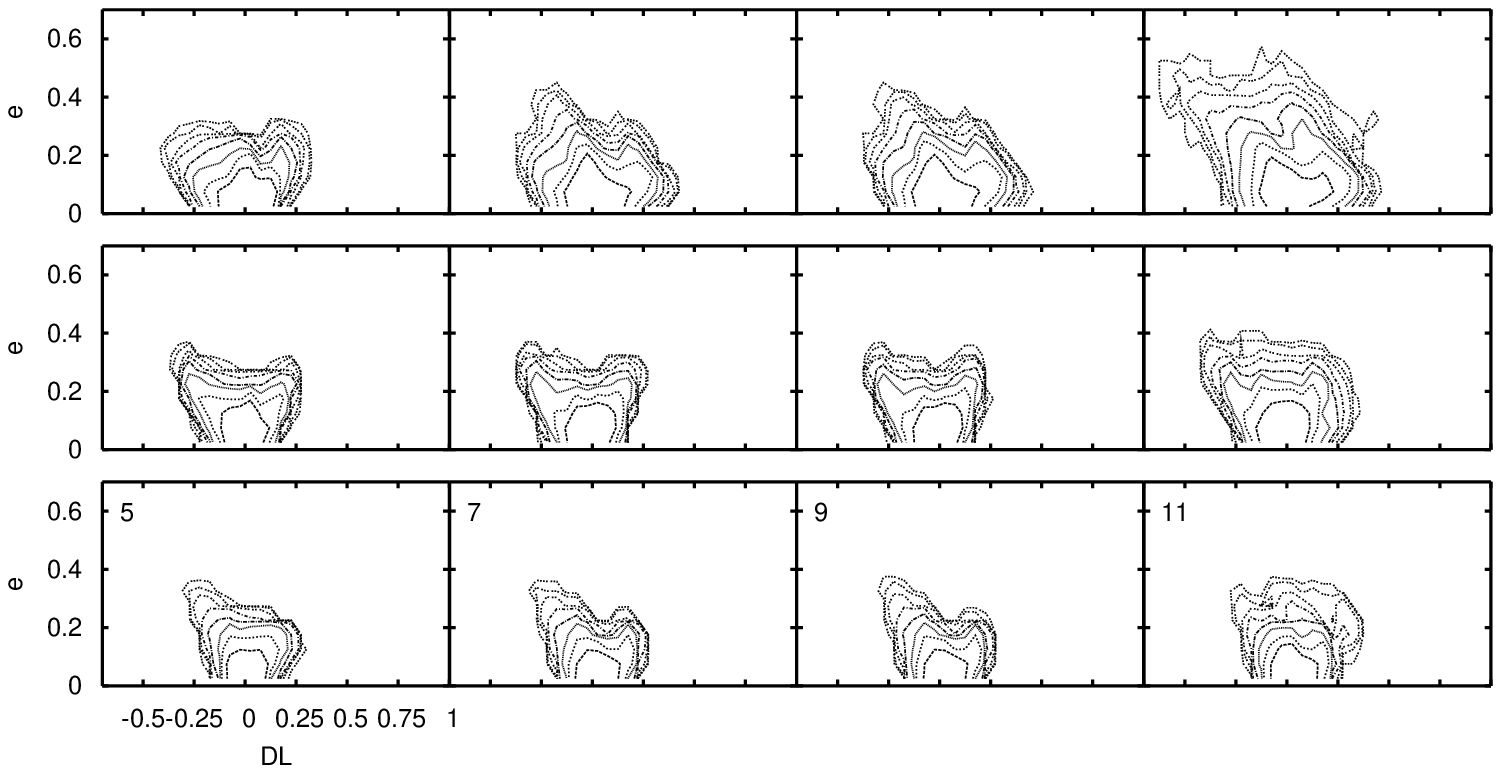}
\includegraphics[angle=0,width=5.0in]{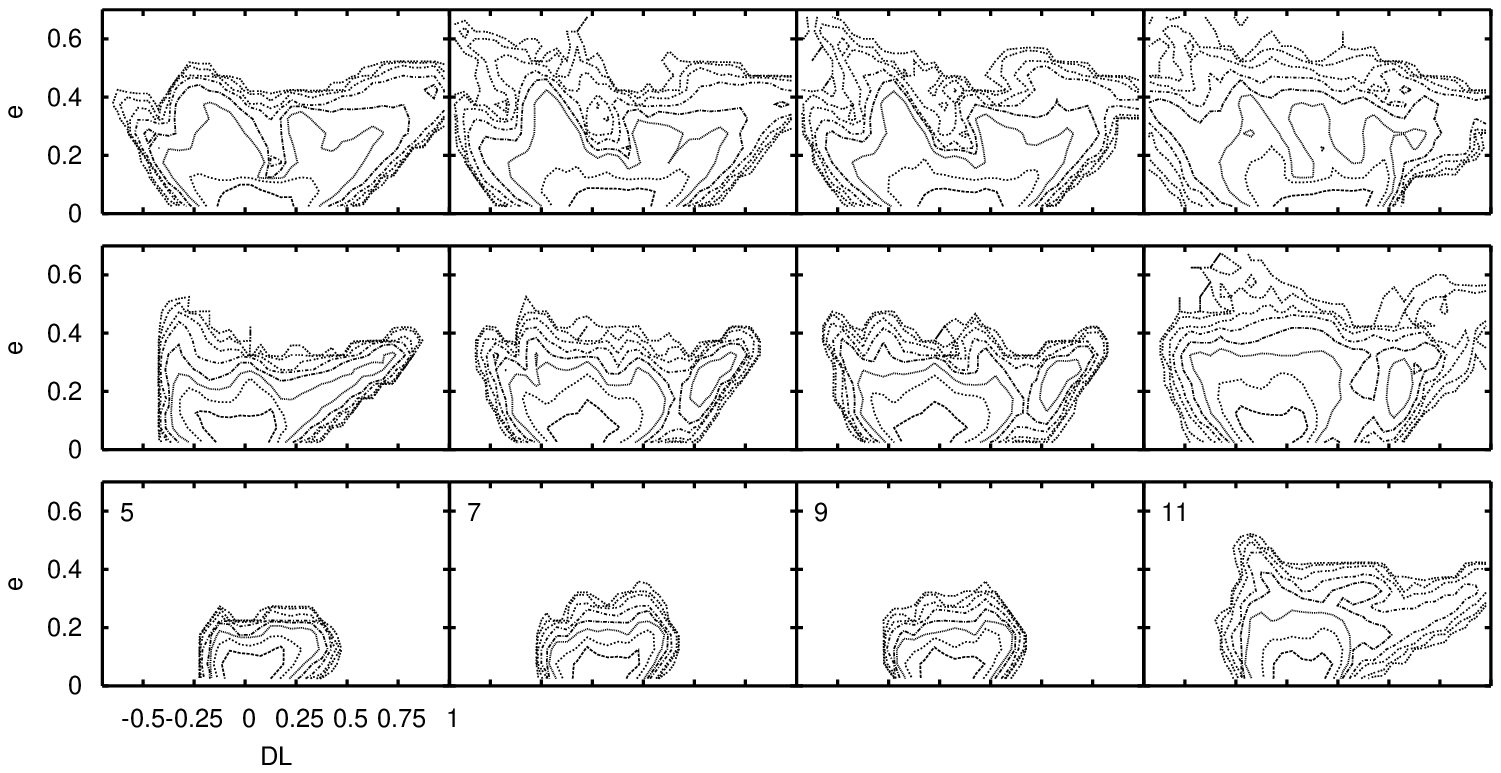}
\caption{Shown is eccentricity versus angular momentum change
distribution for stars  with different initial radii
(top to bottom) and at different times (left to right).
From left to right we show the distribution at times
separated by 2 orbital periods at $R_0$.  Times
are given on the upper left of the lower set of panels.
The leftmost panels show the distribution when the satellite
at the first apocenter (after one pericenter passage)
and the rightmost panels when it is at its fourth apocenter.
From top to bottom we show the  distribution
for stars with different initial radii.
The top panels show stars with initial radii in the range 
1.3--1.5, the middle panels with initial radii
1.0--1.3 and the bottom panels with radii between 0.8 and 1.0. 
Contours are separated by a factor of $\sqrt{10}$.
Correlations between eccentricity and  angular momentum
change are reduced after a few pericenter passages.
There is a population of low eccentricity stars
with different final angular momentum.
These correspond to stars in nearly circular orbits
that have been moved radially from the radii
of their birth.
The distribution is wide for stars with initial radii near the
the pericenter distance of the satellite orbit.
a) For simulation A1 with satellite on a polar orbit.
a) For simulation A2 with satellite on an inclined orbit.
\label{fig:dle}
}
\end{figure}

In Figure \ref{fig:dle} 
shows eccentricity, $e$, versus angular
momentum change, $dL$, at the times separated by 2 $P_0$
for simulations A1 and A2.
Each row is computed for a different range of initial
radii.  The top row corresponds to initial radius in the range
1.3 --1.5, the middle row 1.0 -- 1.3 and the bottom
row 0.8--1.0.  The left hand panels show
the $e$ versus $dL$ distribution 
following only a single pericenter passages.
At this time little heating has taken place and there
are strong correlations between angular momentum change
and particle eccentricity caused by the satellite.  However
after 3 more pericenter passages the correlations
between these properties are much weaker. 
From the top-right panels,  particularly in Figure \ref{fig:dle}b
for the inclined satellite orbit,
we can see that stars in
the outer disk can experience changes in their angular
momentum but also be placed in low eccentricity orbits.  
This means that a particle in a circular orbit
can be moved to a nearly circular orbit
with a different radius.  This change can be called radial
migration though here it is caused by gravitational scattering.

Stars with negative $dL$ have orbits with mean
radii interior to their initial radius 
whereas stars with positive $dL$ have mean radii
outside their initial radii.
We see from Figure \ref{fig:dle}b
that a significant population of stars from the outer galaxy
that crosses into the inner galaxy.
This implies that perturbations from
satellites or subhalos can cause stars from the outer disk
to reach the solar neighborhood.
This mechanism provides a possible explanation for
the population of moderate metallicity disk stars
with positive $v$ velocity components originating
in the outer galaxy that are present in the solar neighborhood
and are discussed by \citet{haywood08}.

We see in Figure \ref{fig:dle}b that there are also
moderate eccentricity 
stars with positive $dL$ and so 
mean radii exterior to their initial radius.
This population of stars forms a diffuse extended stellar disk 
possibly similar to those sometimes  seen
seen in the outskirts of other galaxies \citep{bland05}.
Because of the large wide distribution of $dL$ and $e$
at later times in the outer disk (Figure \ref{fig:dle}b)
the outer disk would have a reduced metallicity gradient
but large metallicity scatter compared to the inner disk.
Perturbations from satellites
could account for the flattening of the metallicity gradient in
the outskirts of other galaxies \citep{vlajic09}.

While we have considered only one satellite here,
\citet{kazantzidis08} suggest that
a subhalo of this size can be expected to interact with a Milky Way
sized galaxy approximately every Gyr.
A galaxy of this size is suspected to have left
behind $\omega$ Cen and the 
$\omega$ Cen moving group \citep{bekki03,meza05}.
As the orbit chosen here was short, significant disk heating 
occurred in a short timescale ($\sim 2.5$ Gyrs).
Future studies could consider heating and mixing by
more than one satellite (or halo substructure), those that
merge and those drawn from a distribution consistent
with hierarchical galaxy formation models.

\begin{figure}
\includegraphics[angle=0,width=4.2in]{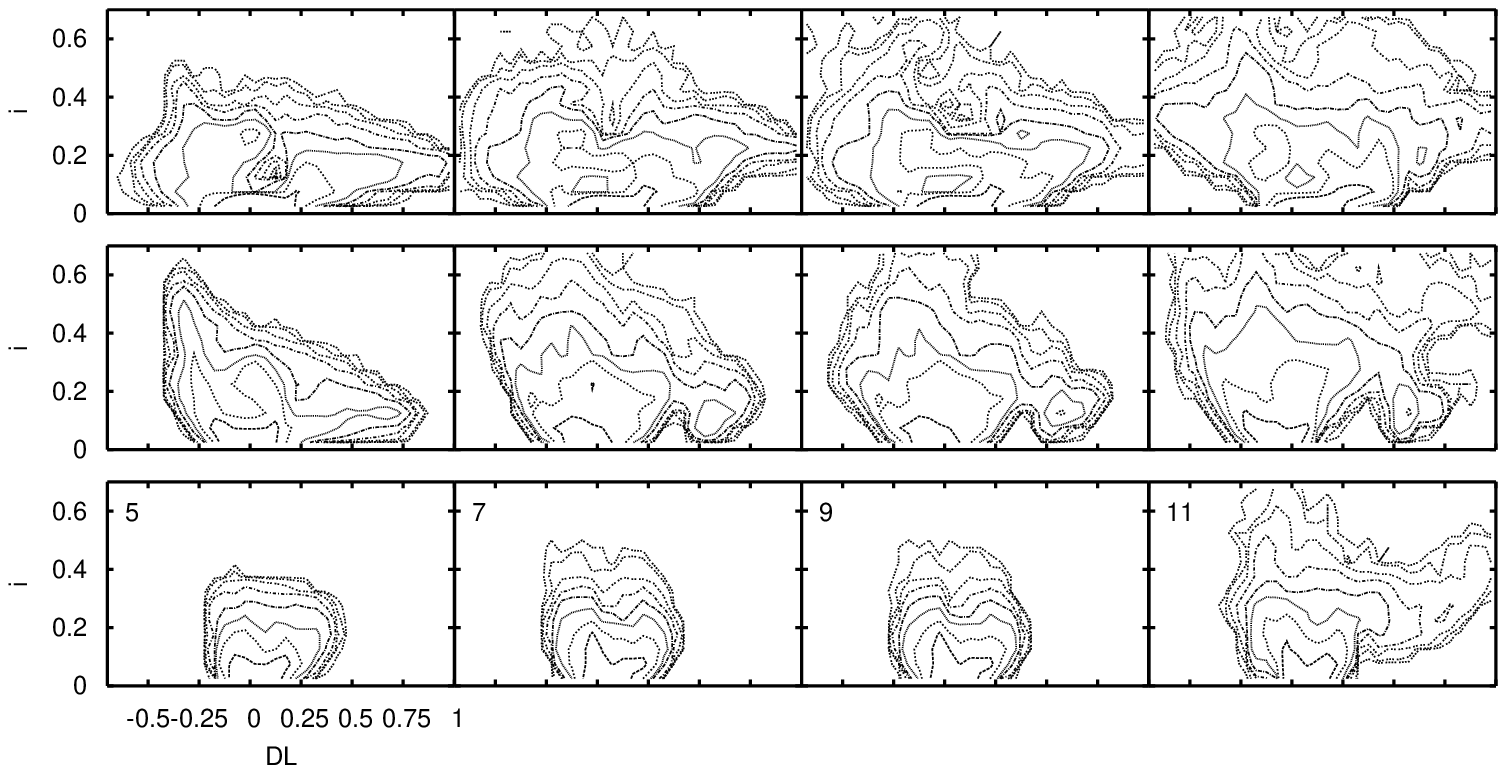}
\includegraphics[angle=0,width=4.2in]{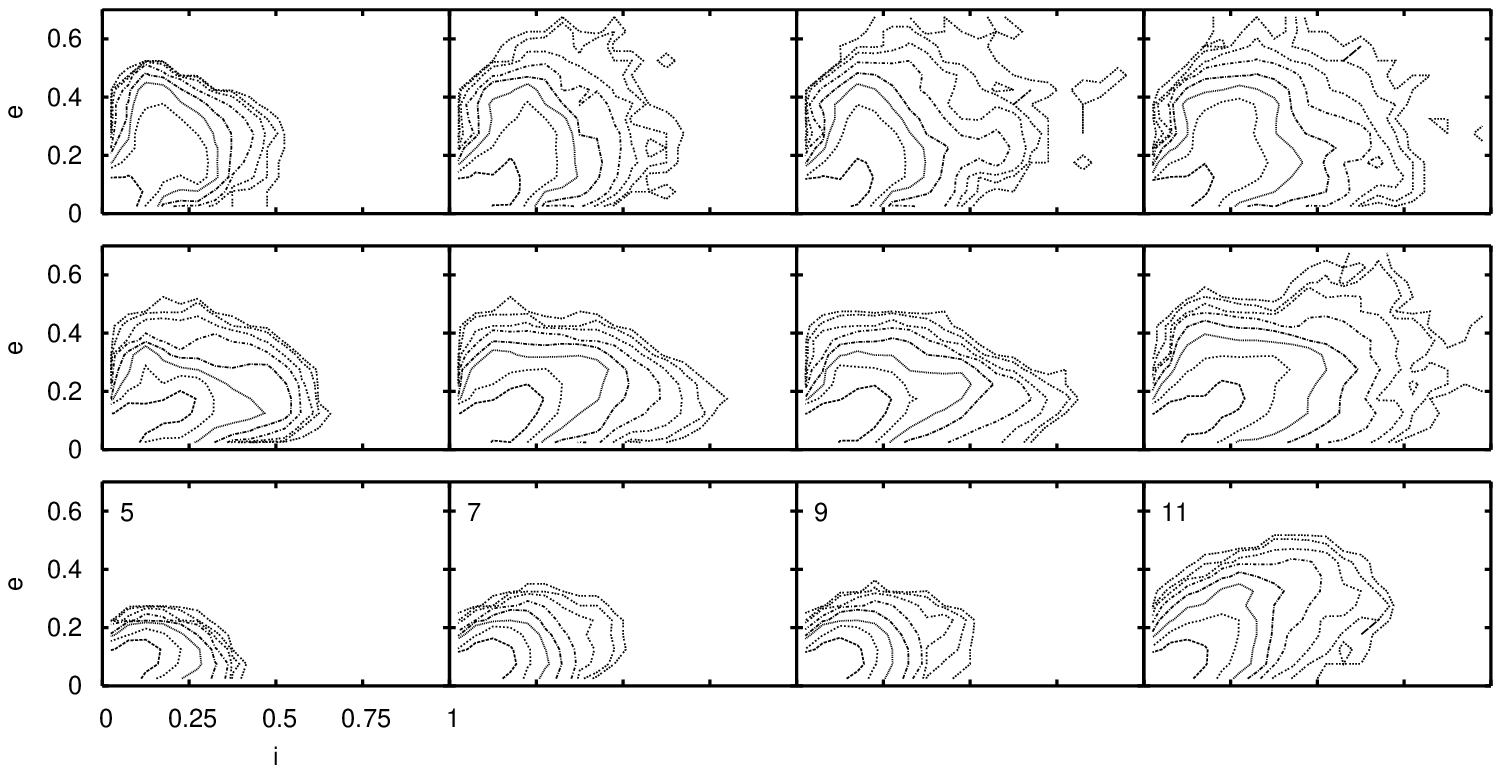}
\caption{a) Inclination versus angular
momentum distribution for simulation A2.
This is displayed similarly to 
Figure \ref{fig:dle}.
b) Inclination versus eccentricity distribution
for simulation A2.
We find that correlations between eccentricity and
inclination are also reduced after a few pericenter passages.
\label{fig:dli}
}
\end{figure}

In Figure \ref{fig:dli} we show the distribution
of angular momentum change versus inclination
and the eccentricity versus inclination change, for
stars with different initial radial bins.
The inclination for each star is estimated
as $i \approx \sqrt{v_z^2 + z^2/r^2}$.
Correlations between eccentricity and inclination 
(see Figure \ref{fig:dli}b)
and between inclination and angular momentum change
are also much reduced after a few satellite orbits.
Stars moved into the inner galaxy (the left hand 
side of panels in Figure \ref{fig:dli}a can also
be on moderate inclination orbits.  We note that
there is a deficit of low eccentricity and low inclination
stars that have experienced large angular momentum
changes, so correlations between these quantities have
not been completely erased at later times.

\section{Discussion}

In this paper we have used particle integration simulations
to investigate the effect on the Milky Way disk
of a fairly low mass satellite in
a tight orbit around the Milky Way.
We find that a warp, lopsidedness and spiral structure
are excited in the outer disk by a satellite in an orbit similar
to that estimated by the Sgr dwarf galaxy.
These structures
are strong providing the mass of the satellite is 
$\ga  5 \times 10^9 M_\odot$, the orbit pericenter is small,
$\la 10$~kpc.  Perturbations are stronger when the satellite
is in an inclined rather than polar orbit.
If the orbital period of the satellite is short
compared to the relaxation timescale of the disk
then structure is increasingly excited each
orbit and the disk continuously displays structure such
as lopsidedness, a warp and spiral structure.
Stellar orbits can intersect implying that 
the stellar distribution and gas distribution in 
the outer Galaxy could differ following tidal perturbations
by a satellite.  It is interesting to consider the possibility
that differences between
the stellar and gas distribution might
be used to differentiate between models with and without
warp-mode and lopsided-mode excitation. 

Perturbations in the disk are also seen in 
a constructed velocity distribution at a specific location
in the disk.  This procedure creates a simulated version of
the velocity distribution in the Solar neighborhood.
Strong low velocity streams are present within $40$km/s of
the local standard of rest.  Smaller
streams at higher velocities can also be present, mostly
at positive $v$, representing stars that come into
the solar neighborhood from the outer galaxy. 
We infer that moderate velocity 
streams in the solar neighborhood velocity
distribution could be
caused by strong recent tidal perturbations in the outer Galaxy.

We find that correlations between eccentricity, inclination and 
and angular momentum change are reduced
after a few pericenter satellite passages.  The eccentricity
and angular momentum distribution is broad for
stars born near the satellite pericenter radius,
and includes low eccentricity stars.
Correlations between inclination and eccentricity
are reduced after a few satellite orbits.
Disk stars born in the outer galaxy can
come into the inner galaxy. Some of them
can be put in low eccentricity orbits at
radii distant from their birth places.
We infer that radial mixing and migration
can be induced by satellite or subhalo perturbations 
in the outer disk.  Such a scenario may
account for the lack of strong correlation between
metallicity and space velocities and
moderate metallicity stars at positive $v$ 
seen in the Solar neighborhood. 
Such a scenario may also account for flattening of the metallicity
gradient in the outskirts of other galaxies \citep{vlajic09}.

As the simulations here integrated test particle
trajectories in an isothermal gravitational potential 
they only poorly approximate the dynamics
of the Galactic disk.  The simulations here could 
be redone with N-body simulations and more realistic
galactic mass models.
Simulations which include a self-gravitating disk can investigate
the excitation of global modes (e.g., \citealt{jog08}) 
in context with disk heating, radial mixing and the structure
of the stellar velocity distribution.
The satellite orbit chosen for this study is extreme
in that the pericenter distance from the Galactic center
is small and the orbital period short (as is true
for the estimated orbit of Sgr dwarf galaxy and that estimated
for $\omega$~Cen's progenitor; \citealt{bekki03}).
We did not allow the satellite orbit to decay 
via dynamical friction, be tidally stripped or merge with the Milky Way.  
As the satellite spirals inwards due to dynamical friction its
mass would be reduced as it is tidally stripped by the Milky Way 
(e.g., \citealt{zhao04}).
However it may continue to perturb the Galaxy disk as each
pericenter passage brings it increasingly inward. 
A satellite with a higher central concentration would 
not be tidally disrupted until
the satellite reaches a smaller galactocentric radius
and so could perturb the disk over a larger range of radius
than a diffuse object that is disrupted in the outer halo.
A more realistic N-body based simulation that
can exhibit variations in the Galactic potential and satellite orbit 
would probably show increased heating and migration
but weaker correlations between disk
stellar eccentricity, inclination and distance from birth location
than the simulations explored here.
Future studies could consider more realistic satellite orbits, 
satellites that merge with the Galaxy and more than one satellite. 

We have demonstrated that radial mixing can be
caused by tidal perturbations from satellite galaxies 
but much work remains to quantify the extent of the mixing 
and determine how it depends on satellite mass and orbit.
Future work could compare the importance of mixing
caused by satellites and that caused by 
transient spiral structure.
If mixing associated with
tidal perturbations from satellite galaxies 
dominates and is required to account for
metallicity and velocity distributions in some regions
of the Milky Way disk then one might place statistical 
constraints on the number, mass and orbits of objects
that have perturbed and merged with the Milky Way from 
the observed distributions.
While we have seen correlations between eccentricity,
inclination and angular momentum change reduced
after a few satellite pericenter passages, weak correlations between
these quantities do persist.    Futures studies should
continue to search for correlations in these quantities
to probe both merger remnant related  and tidally induced
structure models.

\vskip 0.2 truein
We thank Richard Edgar for invaluable help with the GPU cluster.
We thank NVIDIA for giving us two graphics cards.
We thank Eric Mamajek and Jeremy Bailin
for helpful discussions.
JBH is supported by a Federation Fellowship
from the Australian Research Council.
Support for this work was provided by NASA through an award
issued by JPL/Caltech,
by NSF grants AST-0406823 \& PHY-0552695 and
and HST-AR-10972 to the Space Telescope Science Institute.
This work is based on observations made with the Spitzer Space Telescope, which is operated
by the Jet Propulsion Laboratory, California Institute of Technology under a contract with NASA.

{}

\begin{table*}
\begin{minipage}{120mm}
\caption{Test Particle Simulations \label{tab:tab1}}
\label{tab:single}
\begin{tabular}{@{}lc}
\hline
Parameter & Value \\
\hline
$R_{apo}$  & 6.0$R_0$ \\  
$R_{peri}$ & 1.3$R_0$  \\  
$P_R$ & 2.8  $P_0$ \\  
$M_p$ & 0.06 $M_0$ \\
$a$ & 0.1 $R_0$ \\
$q$ & 0.95  \\
\hline
\end{tabular}
{\\
$R_{apo}$ and $R_{peri}$ are the approximate 
galactocentric radii of apocenter pericenter in units of $R_0$.
These are not exact as the halo is mildly oblate.
$M_p$ is the mass of the perturber in units
of $M_0$ which is $\sim 10^{11} M_\odot$.
$P_R$ is the period of radial oscillations of the satellite's orbit
in units of that of the orbital period of Sun around the Galaxy.
Initial disk particles are chosen between 0.5 and 2.5 $R_0$.
The parameter $a$ is the softening length of the satellite
and the core radius of the rotation curve. 
The parameter $q$ allows the halo to be mildly oblate.
Simulation A1 has satellite in a polar orbit.
Simulation A2 has satellite in a prograde orbit with spin inclination
30$^\circ$ from the Galactic pole.
The simulations begin with satellite at apocenter.
}
\end{minipage}
\end{table*}


\clearpage
\clearpage

\clearpage

\clearpage

\clearpage

\end{document}